\documentclass{ws-p10x7}
\usepackage{xspace}
\usepackage{amssymb}
\newcommand{\photon}{\ensuremath{\gamma}\xspace}
\newcommand{\fgamma}{\ensuremath{F_2^{\gamma}}\xspace}
\newcommand{\qsq}{\ensuremath{Q^2}\xspace}
\newcommand{\gev}{\mbox{\rm ~GeV}\xspace}
\newcommand{\gevsq}{\ensuremath{{\rm ~GeV}^2}\xspace}
\newcommand{\ar}{\rightarrow}
\newcommand{\itemp}{\protect\vspace*{-0.2cm}\item}
\newcommand{\epem}{\ensuremath{e^+e^-}\xspace}
\newcommand{\ppbar}{\ensuremath{p\bar{p}}\xspace}
\newcommand{\xpom}{\ensuremath{x_{_{I\!\!P}}}\xspace}
\newcommand{\pom}{\ensuremath{I\!\!P}\xspace}
\newcommand{\fdfour}{\ensuremath{f_i^D(x,\qsq,\xpom,t)}\xspace}
\newcommand{\fdthree}{\ensuremath{F_2^{D(3)}(\beta,\qsq,\xpom)}\xspace}
\newcommand{\fdtwo}{\ensuremath{F_2^{D}(\beta,\qsq)}\xspace}

\newcommand{\these}{ these proceedings}

\newcommand{\coll}[1]{#1 Coll., }
\newcommand{\collh}{\coll{H1}}
\newcommand{\collz}{\coll{ZEUS}}
\newcommand{\collc}{\coll{CDF}}
\newcommand{\colld}{\coll{D0}}
\newcommand{\collo}{\coll{OPAL}}
\newcommand{\colll}{\coll{L3}}

\newcommand{\journal}[4]{ {\it #1} {\bf #2} (#3) #4}
\newcommand{\nuclphys}[3]{\journal{Nucl.Phys.}{#1}{#2}{#3}}

\newcommand{\physlet}[3]{\journal{Phys.Lett.}{#1}{#2}{#3}}
\newcommand{\physrev}[3]{\journal{Phys.Rev.}{#1}{#2}{#3}}
\newcommand{\prl}[3]{\journal{Phys.Rev.Lett.}{#1}{#2}{#3}}
\newcommand{\epj}[3]{\journal{Eur.Phys.J.}{#1}{#2}{#3}}
\begin{document}

\title{Soft Hadronic Interactions}
\author{Peter Schleper}
\address{DESY, Notkestr. 85, 22607 Hamburg, Germany\\E-mail: Peter.Schleper@desy.de}

\twocolumn
[\maketitle\abstract{Recent developments in soft hadronic interactions are
reviewed. 
Emphasis it put on measurements of the proton structure at
low $x$, photon structure, diffraction and exclusive processes such
as vector-meson production and their interpretation in approaches to
QCD dynamics like BFKL or CCFM.}] 
\section{Introduction}
Quantum Chromodynamics is the generally accepted field
theoretical prescription of strong interactions and is successfully
applied to processes where a hard scale is
present, given by either a 
highly virtual particle, a large transverse momentum or a large mass
of the exchanged particles\protect\cite{nania}.
In such processes the strong coupling constant is small enough to
allow for perturbative calculations and QCD is predictive. 
In soft processes, where such a hard scale is not present,
$\alpha_s$ becomes large and perturbative techniques are not
applicable. This prohibits predictions of  such fundamental quantities
as the size and mass of the proton, the total cross-section for
hadron-hadron scattering or cross-sections for elastic scattering
of hadrons. All these questions are closely
connected to confinement and still belong to the least
well understood properties of strong interactions.

In the past many phenomenological models have been developed to
describe soft interactions.
More recently, driven by new data obtained in the transition
region between soft and hard processes from the HERA, LEP and Tevatron
experiments, the theoretical interpretation within perturbative QCD
(pQCD) has made
considerable progress. This is also the focus of
this review\footnote{Plenary Talk, ICHEP 2000, Osaka.},
and hardly any reference is given to phenomenological prescriptions of
soft physics.
\subsection*{Deep Inelastic Scattering}
\begin{figure}[bthp]
\epsfxsize0.5\linewidth
\figurebox{}{}{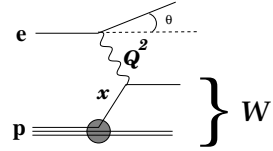}
\caption{Feynman diagram for deep inelastic scattering.}
\label{feyn.dis}
\end{figure}

\begin{figure}[bthp]%
\epsfxsize\linewidth
\figurebox{}{}{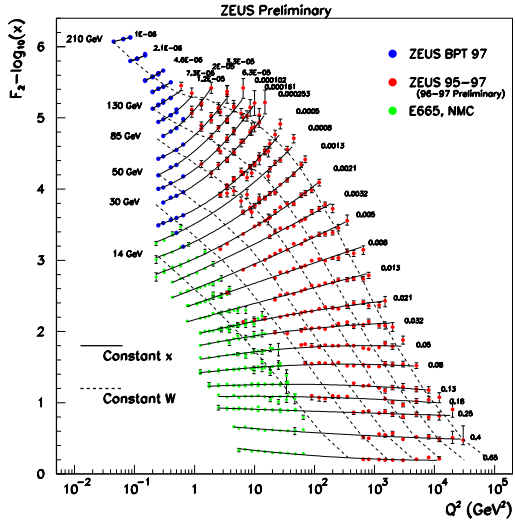}
\caption{Proton structure function $F_2$ as a function of \qsq with
lines of constant $x$ and $W$ from \protect\cite{pellegrino}.  
}
\label{f2.alldata.zeus}
\end{figure}

To introduce the concepts it is useful to start with the example of
deep inelastic scattering (DIS) of a lepton on a proton (Fig.~\ref{feyn.dis}).
Here $Q^2$ denotes the virtuality of the exchanged photon $\gamma^*$, 
the Bjorken scaling variable $x$ corresponds in lowest 
order to the momentum fraction of the struck quark in the proton and
$W$ is the total centre of mass energy in the $\gamma^*$-proton system.
For $Q^2\gg\Lambda_{QCD}^2$ the high $\gamma^*$ virtuality provides the
hard scale and the structure of the proton is resolved into partons,
i.e. the cross-section is proportional to the structure
function $F_2(x,\qsq)$ which measures the quark momentum distribution in the
proton. Fig.~\ref{f2.alldata.zeus} shows the data on  $F_2$  as
obtained by fixed target experiments and at HERA\cite{pellegrino},
which now cover a \qsq range
from several $10^4$\gevsq down to $\simeq \Lambda_{QCD}^2$,
and include momentum fractions $x$ as low as $10^{-6}$. 
The data nicely demonstrate the point-like nature of quarks in the region of
approximate scaling at $x\approx 0.1$, modified by negative scaling
violations at 
higher $x$ which are attributed to the quark splitting $q\ar qg$ and,
at low $x$, 
by positive scaling violation due to the prevailing gluon splitting
$g\ar gg,\,q\bar{q}$. 
The sharp increase of $F_2$ with decreasing $x$ at \qsq values of a
few \gevsq, the main result of the first HERA data \cite{first.hera.f2}, is of
great importance as this slope is directly related to the gluon density in
the proton. 
The proton at low $x$ is thus a system of very high
gluon density and thereby a unique environment for the understanding
of QCD dynamics. 
 Since
$W^2=Q^2(1-x)/x$, the extrapolation towards low $x$  at fixed \qsq
corresponds to the high energy ($W$) limit of QCD, which is
interesting in itself but also has impact for e.g. cosmic ray
experiments, heavy ion collisions and 
Higgs production at the LHC via $gg\ar H$ \cite{lhc}.
\subsection*{QCD Dynamics at Low $x$}
\begin{figure}[bthp]  
\begin{center}
   \epsfig{file=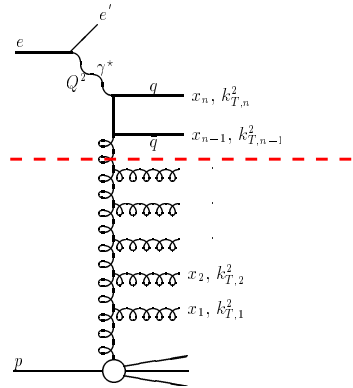,width=0.7\linewidth}
   \caption{\label{gluonladder} The parton ladder in a hard scattering
   process. The dashed line denotes the factorisation into a hard
   matrix element and the parton density.} 
\end{center}
\end{figure}

To better understand the present excitement about low $x$ dynamics it
is worth recalling the assumption of {\it
factorisation} of hadronic cross-sections into a matrix element of
the hard process (which is calculable in pQCD) and parton
distribution functions 
which represent all other soft parts of the diagrams 
(Fig.~\ref{gluonladder}). 
In the standard Altarelli-Parisi (DGLAP\cite{DGLAP}) evolution the phase-space 
for parton emission is approximated by summing those
contributions where the transverse momenta $k_{T,i}^2$ along the
ladder strictly increase from the proton to the hard scattering, and
therefore is applicable only at larger \qsq.
To this approximation the process
factorises into calculable coefficient functions and supposedly
universal parton distribution functions which obey the DGLAP evolution
equations.

It is apparent that other momentum configurations will contribute both
at  low \qsq and at low $x$. At low \qsq (i.e. \qsq  not much larger
than $\Lambda_{QCD}^2$) at any point along the ladder 
$k_{T,i}^2$ might exceed \qsq which
destroys the required strong $k_T$ ordering. Even at higher \qsq but
low $x_n$, the large possible differences between the longitudinal
momenta $x_i$ imply sufficient phase-space also for large
transverse momenta $k_T$ somewhere along the ladder, again affecting 
strong $k_T$ ordering.
The transition region from high to low \qsq and the low $x$ limit 
thereby elucidate other approaches to pQCD and hence to
QCD dynamics.

If the strong $k_T$ ordering criterion is relaxed the approximations
become more 
complex. For the BFKL\cite{BFKL} and CCFM\cite{CCFM} evolution
equations ordering in $x$ or 
in the emission angle of partons along the ladder is assumed,
respectively.
The resulting ``unintegrated'' parton densities depend directly on the
transverse momenta $k_T$  as shown in Tab.~\ref{QCDevolution}.
\begin{table}[bthp] 
\begin{center}
\caption{\label{QCDevolution} QCD evolution equations based on
         different ordering schemes, the dependence of the parton
         density functions (pdf), the terms summed and the kinematic
         range of application.}  
\begin{tabular}{|l|l|l|l|} \hline
             & DGLAP       &BFKL                 & CCFM      \\ \hline
    order    & $k_T$       & $x$                 & angle     \\
    pdf      & $f(x,Q^2)$  & $f(x, k_T^2)$       & $f(x, k_T^2, Q^2)$ \\
    $\sum$   & $\ln Q^2$   & $\ln 1/x$           & $\ln Q^2$ +  \\  
             &             &                     & $\ln 1/x$ \\
    valid    & high \qsq   & $k_T^2 \approx Q^2$ & low $x$ \\
             & high $x$    & low $x$             &  \\ \hline
\end{tabular}
\end{center}
\end{table}
The BFKL approximation is expected to be valid only at low $x$, since
it does not contain the DGLAP terms.
It is not known yet how low \qsq or $x$ have to be to yield sizeable 
BFKL type contributions.
The CCFM equation would in principle enable a smooth
extrapolation between the DGLAP and the BFKL regime as it contains
both parts. Up to now it is
only applicable at low $x$ since the quark splitting terms are not
known yet.
\subsection*{Vector-Meson Production}
\begin{figure}[h]%
\epsfxsize0.7\linewidth
\figurebox{}{}{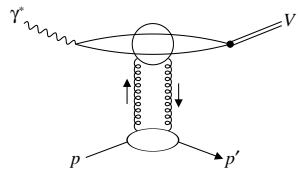}
\caption{Feynman diagram for the production of
   vector-mesons $V$ on an elastically scattered proton via 2-gluon exchange.}
\label{feyn.vm}
\end{figure}

An intriguing view of the interplay between soft and hard physics is
derived from {\it elastic photoproduction} ($\qsq\approx 0$) of
vector-mesons at HERA\cite{mellado}, $\gamma p \ar V p$.
\begin{figure}[bth]
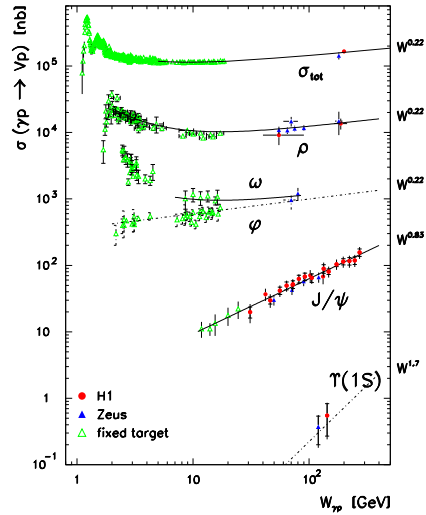
%
\epsfxsize0.8\linewidth
\figurebox{}{}{vmall_col.epsi}
\caption{Energy dependence of the total
   photon-proton cross-section $\sigma_{tot}$ in comparison to the
   cross-section for elastic photoproduction of vector-mesons 
   from HERA \protect\cite{mellado}.} 
\label{vm.all}
\end{figure}
For vector-mesons consisting of light quarks ($\rho, \omega$) the
energy dependence of the cross-section $\sigma_{\gamma p\ar Vp}$ is
very weak and similar to the total photon-proton cross-section
(Fig.~\ref{vm.all}). For the $J/\Psi$, however, the energy dependence is 
significantly stronger, indicating that the charm quark mass provides a
hard scale. The size of the cross-section is hence not of
geometrical nature but can be associated with the partonic content of
the proton. In this case the process is assumed to be dominated by the
exchange of a pair of gluons which together form a colour singlet to yield an
elastically scattered proton (Fig.~\ref{feyn.vm}).
The increase of the cross-section towards large $W$ reflects
the increase of the  
gluon density (squared) towards low $x$. 

It is evident that in the extreme high energy limit
this behaviour must change since the $J/\Psi$ contribution should
never exceed the total cross-section. New dynamics must therefore
dampen the  $J/\Psi$ cross-section in the high $W$ (low $x$) limit.

In QCD calculations\cite{jpsi.qcd} the cross-section is explained as a
three step 
process: the splitting of the photon into a $q\bar{q}$ dipole, the
interaction of this dipole with the gluon pair and finally the
formation of the vector-meson. At low \qsq and low $x$ the three steps
take place on very different time scales suggesting that the cross-section
factorises into the probabilities for each of the individual steps.

This factorisation allows the same dipole cross-section to be applied
also in other processes at low $x$ such as inclusive DIS, jet 
production or diffraction\cite{dipolefac,dipolemod}. 

\section{The Proton at Low $\mathbf{x}$}
The proton structure at low $x$ is of interest not only
as a new domain in QCD, where fundamental insight into
the dynamics within and beyond the perturbative regime is still to be
gained. It is also of relevance for the program at the Tevatron and the
LHC\cite{lhc}. 
For the small Higgs mass expected due
to the indirect and direct measurements at LEP \cite{lep.higgs} the
dominant production  
process $gg\ar H$ (with $M_H=x_1\,x_2\,s_{\mathrm{LHC}}$) at LHC
energies implies gluon momenta $x_i$ in the 
range $1>x>10^{-4}$, or $0.1>x>10^{-3}$ if the angular acceptance
for the Higgs decay products is restricted to rapidities $|y|<2$. 
Sensitive tests of the Higgs sector therefore crucially depend on the
gluon density at low $x$ and the reliability of the
theoretical extrapolation from low to high \qsq \cite{lhc}.
At and below $x=10^{-3}$ the only process giving access to the gluon
density in the proton is DIS at HERA.
\subsection*{HERA Data and the Gluon Density}
\begin{figure}[bthp]
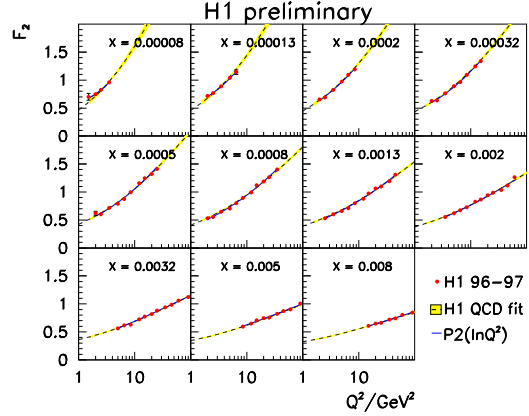

\epsfxsize\linewidth
\figurebox{}{clip=}{H1prelim-00-141.fig9.epsi}
\caption{The proton structure function $F_2$ at low $x$ as a function of
         \qsq from H1 \protect\cite{h1.f2,h1.qcd}. 
}
\label{h1.f2}
\end{figure}

Both HERA collaborations H1 and ZEUS have released new preliminary
data\cite{pellegrino,h1.f2} on 
their structure function measurements in the low $x$ and low \qsq
region. The H1 data\cite{h1.f2} shown in Fig.~\ref{h1.f2} have now in a
large range 
a statistical precision of $\approx 1\%$ and 
systematic errors of about $3\%$ due to new instrumentation such
as silicon tracking and high granularity calorimetry. 
\begin{figure}[bthp]
\epsfxsize0.7\linewidth
\figurebox{}{clip=}{H1prelim-00-141.fig16b.epsi}
\caption{The gluon density obtained by H1 \protect\cite{h1.qcd} in a
   DGLAP fit 
for different minimal $Q^2_{min}$ values of the
   input data.}
\label{h1.gluonfit}
\end{figure}
\begin{figure}[bthp]
\epsfxsize0.8\linewidth
\figurebox{}{clip=}{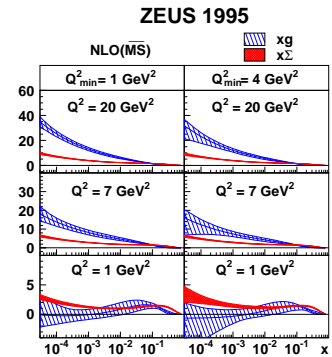}
\caption{The gluon density ($xg$) and quark singlet density ($x\Sigma$)
   at medium and low \qsq from ZEUS\protect\cite{valencegluon.zeus}.}
\label{zeus.gluon}
\end{figure}
This represents important progress in comparison to existing published
data, and the precision is probably close to the final results
attainable by the experiments in parts of the phase-space.
To determine simultaneously $\alpha_s$ and the gluon density at low $x$, the
H1 collaboration has subjected this data to an elaborate NLO QCD
fit taking into account all experimental and theoretical systematic
uncertainties \cite{h1.qcd}.  
A value of  
$\alpha_s(M_Z^2)=0.1150\pm0.0017(stat.)\pm 0.0011(model)\pm
0.005(scale)$ was obtained. The largest uncertainty, arising from the
choice of the renormalisation and factorisation scale, is expected to
be reduced 
significantly once NNLO calculations become available\cite{NNLO}. 
The gluon density is constrained to $\approx 3\%$ 
for $10^{-3}<x<0.1$ at $\qsq=20\gevsq$. 
The fit describes the data well\cite{h1.qcd} down to $\qsq\gtrsim
1\gevsq$, but below $\qsq=5\gevsq$ the resulting gluon distribution
becomes sensitive to the inclusion of data at smaller and smaller \qsq
(Fig.~\ref{h1.gluonfit}). 
This more precise result confirms the previous
finding \cite{valencegluon.h1,valencegluon.zeus} 
that the role of the gluon and quark density change when going to low
values of \qsq (Fig.~\ref{zeus.gluon}). 
At low $x$ the gluon density dominates the 
proton structure for $\qsq \gg 1\gevsq$, but tends to vanish at
$\qsq\approx 1\gevsq$. 
The parton density functions appear to be 
flexible enough for the NLO DGLAP fit to accommodate the inclusive
$F_2$ data down to $\qsq\approx 1\gev^2$, and also describe the 
longitudinal structure function $F_L$ \cite{h1.qcd}. 
Nevertheless this ``valence''-like behaviour of the gluon density
is likely to signal the limit of applicability of the perturbative series in
DGLAP. It might imply that higher twist effects or new dynamics such
as described by the BFKL equation become sizeable. More exclusive data
would be highly welcome in this low $x$ region~\cite{lowx.exclusive}. 
\subsection*{The Low $x$ and Low \qsq Limit}
\begin{figure}[bthp]   
   \epsfig{file=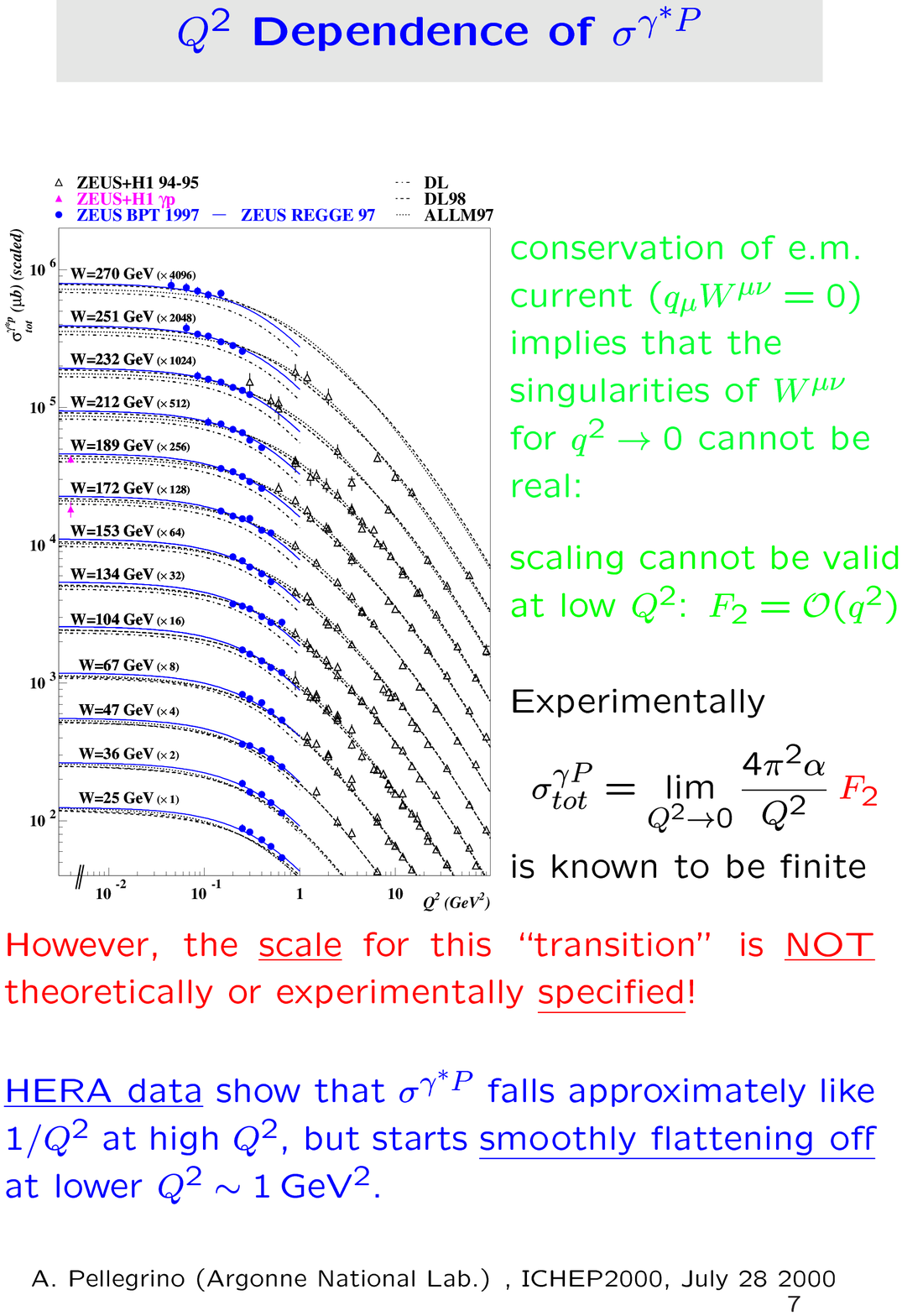,width=0.8\linewidth,
   bbllx=46pt,bblly=252pt,bburx=334pt,bbury=695pt,clip=}
   \caption{\label{sigmatot} The total cross-section for
   $\gamma^*$-proton scattering as a function of \qsq for constant
   values of $W$ from HERA\protect\cite{pellegrino}. Also shown are
   two measurements for $\qsq\approx 0$, from \protect\cite{levonian}.}
\end{figure}

\begin{figure}[bthp]   
   \epsfig{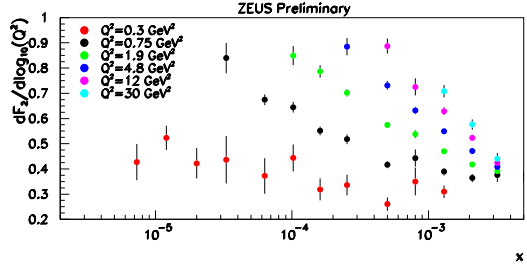}
   \caption{\label{zeus.slopes} The derivative 
   $\left(\partial F_2/ \partial \log \qsq \right)_{\mathrm{fixed}\, x}$, as
   a function of x in bins of \qsq based on data from fixed target
   experiments and from ZEUS \protect\cite{pellegrino}.} 
\end{figure}

Since the CMS energy at HERA is limited to
$\sqrt{s}=320\gev$ kinematics imply that 
values of $x<10^{-4}$ are accessible only at {\it very} low $\qsq <xs$. 
Both H1 and ZEUS have equipped the region close to the beam axis with small
calorimeters and silicon trackers which are able to measure down to
$x\approx10^{-6}$ for $\qsq\gtrsim \Lambda^2_{QCD}$
(Fig.~\ref{sigmatot}).
New data from ZEUS \cite{pellegrino} complement previous measurements on 
$\sigma_{\gamma^* p}\sim F_2/\qsq$ and allow 
interpolation between the steep fall 
of the cross-section at high \qsq and the photoproduction region at
$\qsq\approx 0$, where the cross-section must become independent of
\qsq because of conservation of the electromagnetic current. In
other terms, the photon in the limit $\qsq\ar 0$ fluctuates into a hadronic
object of similar  size as the proton, such that the ``dipole''-proton
cross-section becomes independent of the exact value of
\qsq\footnote{Unfortunately this constant behaviour of the dipole
cross-section in the limit $\qsq\ar 0$ is
sometimes referred to as ``saturation'', which is not the same as
saturation of parton densities in the proton at higher \qsq due 
to recombination effects.}.
The precise form of $\sigma_{\gamma^*p}$ has been subject to 
considerable discussions\cite{foster,dipolemod} about the onset of 
a possible recombination (or saturation) of the gluon density in this
region of low $x$ and high gluon density. Fig.~\ref{zeus.slopes} shows
slopes 
$\left(\partial F_2/ \partial \log \qsq \right)_{\mathrm{fixed}\, x}$,
which to first 
approximation are directly proportional to the gluon density.
For all \qsq, and especially $\qsq\gtrsim 1\gevsq$ where the picture of
a proton resolved into partons is applicable, this slope rises
linearly towards low $x$.
No deviation from this behaviour 
is visible in the energy range accessible at HERA\footnote{
Note that the data shown here is identical to that in a much debated
figure where the same slope is shown in bins of $W$. Due to the
kinematic relation $W^2=Q^2(1-x)/x$ this  
figure however shows a peak at \qsq values of a few $\gev^2$, which
should not be interpreted as saturation of parton densities.}.
In summary the inclusive HERA data do 
not provide evidence for saturation effects of parton densities for
\qsq above a 
few $\gev^2$. At smaller \qsq where the photon itself develops
hadronic structure, low $x$ effects are difficult (if not impossible)
to disentangle from low \qsq effects.
\section{The Photon as a Hadronic Object}
\begin{figure}[bthp]  
   \epsfig{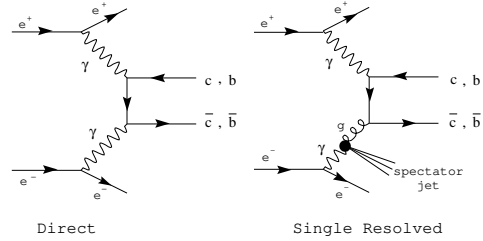}
   \caption{\label{feyn.photon.epem} Example for a direct (left) and single
   resolved (right) $\gamma\gamma\ar \mathrm{hadrons} $ process in
   \epem interactions, shown here for heavy quarks.}
\end{figure}

\begin{figure}[bthp]  
   \epsfig{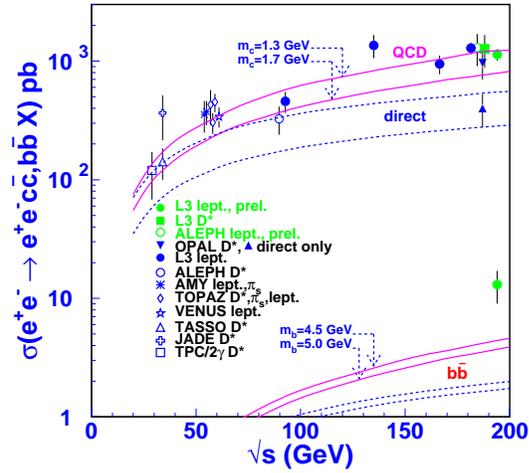}
   \caption{\label{lep.f2gamma.sigma.charm}  The charm and bottom cross-section
   from $\gamma^*\gamma$ scattering at LEP\protect\cite{lep.charm}.}
\end{figure}

\begin{figure}[bthp]  
   \epsfig{file=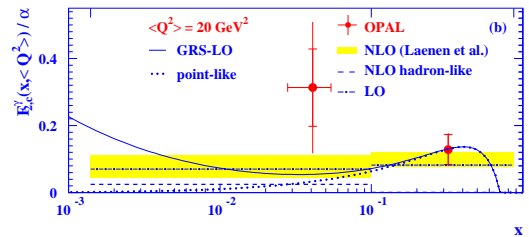,width=\linewidth,clip=}
   \caption{\label{fig.f2gamma.charm}  The charm structure function
   $F_{2,c}^{\gamma}$ from OPAL\protect\cite{opal.f2charm}.}
\end{figure}

A quasi-real photon of very small virtuality $\approx 0$ 
not only couples to other particles directly as a gauge boson 
or as a $q\bar{q}$ dipole of small
transverse size (the point-like component 
which is calculable in pQCD) but can also fluctuate into
a hadron-like object of large transverse size.
The hadronic structure of the
photon can be measured in the processes\cite{nisius,erdmann}:
 \begin{itemize}
    \itemp $e^+e^- \ar e^+e^-\gamma^*\gamma \ar e^+e^- X$ where the
       virtual $\gamma^*$ resolves the structure of the real
       $\gamma$ (Fig.~\ref{feyn.photon.epem}). 
       This process is mainly sensitive to the quark densities in the
       $\gamma$. The gluon density in the photon 
       is only accessible indirectly via the scale dependence which
       requires very precise data. 
    \itemp $ ep \ar e\gamma p \ar e+jets+ X$ at HERA where jets with large
       transverse energy are
       required to resolve the \photon structure. Since coloured
       partons of the 
       proton enter the hard interaction, this process is sensitive
       directly to both the quark and gluon density of the \photon.
\end{itemize}
\begin{figure}[bthp]  
   \epsfig{file=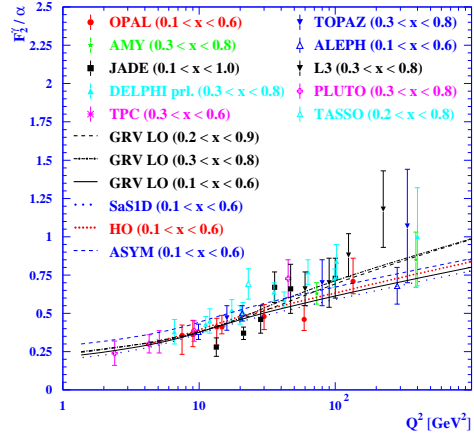,width=0.9\linewidth,clip=}
   \caption{\label{f2gamma.q2} The structure function $F_2^{\gamma}(x,\qsq)$
   for real photons from \epem scattering\protect\cite{nisius,nisiuspriv} as
   a function of the scale 
   \qsq at which the $\gamma$ is probed. Note that \qsq
   here denotes the virtuality of the $\gamma^*$ which
   probes the quasi-real photon of virtuality $\approx 0$.} 
\end{figure}
\begin{figure}[bthp]
   \epsfig{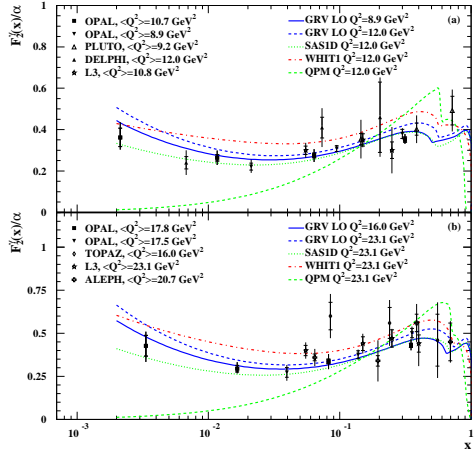}
   \caption{\label{f2gamma.x} The photon structure function
    $F_2^{\gamma}$ from \epem scattering\protect\cite{lep.photon.x} at low
   and high \qsq as  a
   function of $x$. The QPM line which vanishes at low
   $x$ corresponds to the quark-parton model approximation to the
   point-like component.}
\end{figure}
The LEP experiments have recently made substantial progress in the
understanding of both the simulation of the hadronic
final state and the detector response close to the
beam direction. Improved 
unfolding methods have led to much more precise 
measurements and also to better consistency between the experiments.
The LEP II data now are superior to all previous data from e.g. the PETRA
experiments, and in addition give access to
the photon structure at much larger scales and at lower $x$. 
Fig.~\ref{f2gamma.q2} and \ref{f2gamma.x}
show that the photon structure function
\fgamma is now known with a precision of $\approx 10\%$.
The basic expectations for the behaviour of \fgamma are:
\begin{itemize}
   \itemp  \fgamma is dominated at high $x$ by the point-like part.
   \itemp   \fgamma rises with \qsq, in contrast to the proton case, for all
           $x_{\gamma}$ due to the point-like contribution. 
   \itemp  At low $x$ the hadron-like component is expected to dominate and
           the photon becomes similar to the proton, i.e. \fgamma
           rises strongly towards low $x$ and the photon is
           dominated by gluons.
\end{itemize}
The first two points are clearly borne out in the data  shown in 
Fig.~\ref{f2gamma.q2} and \ref{f2gamma.x} when comparing with the expectation
for the point-like component. The hadron-like component is 
seen at low $x$ as the data clearly exceed the point-like part, 
although the expected rise 
of \fgamma at very low $x$ is not significant in the accessible $x$ range.
Note that existing parameterisations of the $\gamma$
structure like GRV(LO) describe
the data well for $\qsq>5\gevsq$. 
Heavy flavour production data from LEP\cite{lep.charm} are shown in
Fig.~\ref{lep.f2gamma.sigma.charm}. 
For charm production the data
agree well with QCD expectations based on the same parameterisations. A
first measurement of the charm structure function $F_{2,c}^{\gamma}$
has been obtained by OPAL\protect\cite{opal.f2charm} (Fig.~\ref{fig.f2gamma.charm}). 
The bottom cross-section as measured by L3
\cite{lep.charm} is 
larger than expected (Fig.~\ref{lep.f2gamma.sigma.charm}), a very
interesting observation as the same trend 
is observed also by the Tevatron experiments and in photoproduction at
HERA. These observations require the theoretical description of 
heavy flavour production processes in hadronic
collisions to be reconsidered.

\begin{figure}[bthp]  
   \epsfig{file=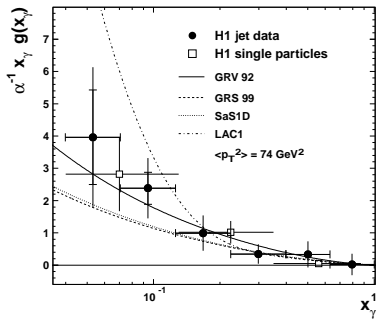,width=0.8\linewidth,clip=}
   \caption{\label{H1.jets.gammap.gluon} The gluon density in the
   photon as measured by H1\protect\cite{H1.jets.gammap}.}
\end{figure}
Further access especially to the gluon component in the $\gamma$ is
obtained from jet production at HERA. In an analysis tailored towards
the low $x$ region, the H1 experiment has used data at relatively low
transverse jet energies $E_T>6\gev$ 
which was corrected for the substantial effects of
secondary interactions between the photon remnant and the proton
remnant in resolved photon processes. 
Subtracting the direct and quark induced parts based on expectation
from $\epem$ data, the gluon distribution in the $\gamma$ is extracted
(Fig.~\ref{H1.jets.gammap.gluon}).  
Albeit only in LO this is the only experimental evidence for a rise of
parton densities in the photon at low $x$.

\begin{figure}[bthp]  
   \epsfig{file=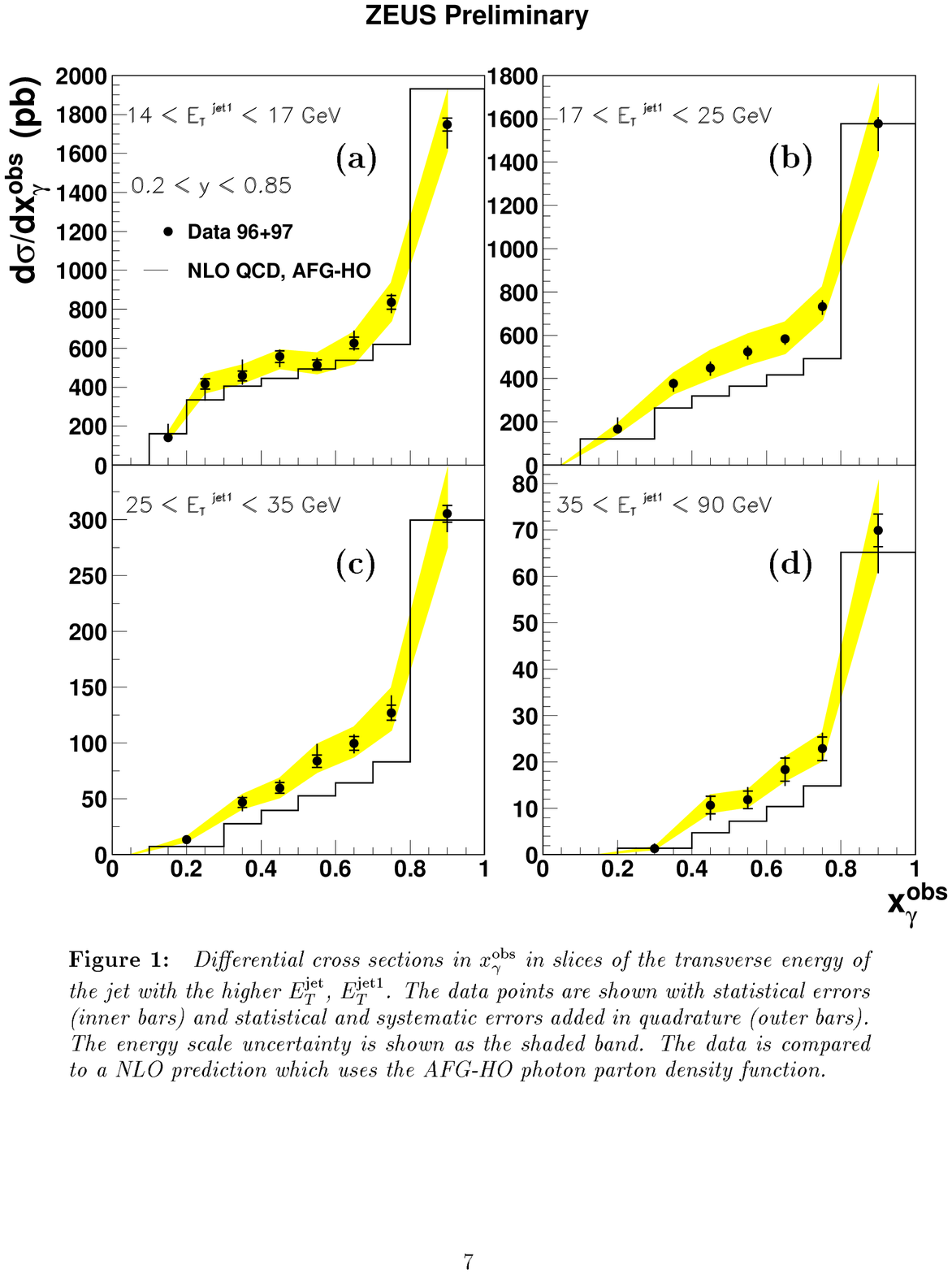,width=\linewidth,
   bbllx=51pt,bblly=280pt,bburx=545pt,bbury=763pt,clip=}
   \caption{\label{Zeus.jets.gammap} Jet cross-section in $\gamma p$
   interactions from ZEUS\protect\cite{Zeus.jets.gammap}.}
\end{figure}
A complementary analysis by ZEUS\cite{Zeus.jets.gammap} uses
jets at much larger $E_T$, which kinematically excludes the low $x$
region. Here the experimental and theoretical uncertainties are
smaller however. In the range $0.3<x< 0.8$ the ZEUS data are
consistent with the NLO calculations for $14<E_T< 17\gev$, but exceed
the calculations for larger $E_T$  (Fig. \ref{Zeus.jets.gammap}).
A similar effect was seen in a first jet measurement in \epem scattering
by OPAL \cite{wengler}. 

Note that this poses a question on the overall consistency of the
photon structure data. The LEP \fgamma data, which are sensitive mainly
to quarks up to scales of $\lesssim 800\gevsq$, as well as the H1 data
on gluons and quarks at low $x$, agree with e.g the GRV
parameterisation. The ZEUS jet data, also sensitive to quarks
and gluons but at high $x$ and high $E_T$ where the point-like
component should dominate and 
little freedom due to the hadron-like component is expected, indicate an
increased scale dependence of the parton densities, an effect which is
difficult to understand. 

The improved precision of the data
in an extended $x$  and \qsq range calls for a
new effort in the understanding of the photon structure in  NLO
QCD. Whether parton densities can be derived which 
consistently describe all data sets within errors remains to be seen.
\section{QCD Dynamics at Low $\mathbf{x}$}
\begin{figure}[th]   
    \begin{center}
    \epsfig{file=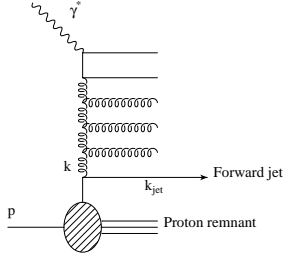,width=0.6\linewidth}
    \caption{\label{feyn.forwardjet} Parton ladder with a hard ``forward'' jet
    close to the proton remnant direction.}
    \end{center}
\end{figure}
In spite of the fact that the inclusive proton structure function
$F_2$ is compatible with the DGLAP evolution equations even at the
lowest accessible $x$ values (for $\qsq\gtrsim 1\gevsq$), it is still
expected that $\ln 1/x$ terms must become 
sizeable in comparison with the $\ln \qsq/\Lambda_{QCD}^2$ terms if only
$x$ is small enough. Considerable effort is therefore expended  at
HERA, as well as at LEP and Tevatron, into the investigation of more exclusive
processes~\cite{bfkl.experimental.evidence,bfkl.theory}. The
advocated\cite{advocate.forward.jets} test case at HERA is 
the production of ``forward jets'' 
(Fig.~\ref{feyn.forwardjet}), i.e. a jet close in rapidity to the
proton. 
In such events, when the jet transverse energy $E_T$
is comparable to the photon virtuality at the other end of the parton
ladder, $k_T$ ordered radiation should be suppressed  and parton
emission might dominate which is  not ordered in $k_T$.

\begin{figure}[h]
  \begin{center}
   \epsfig{file=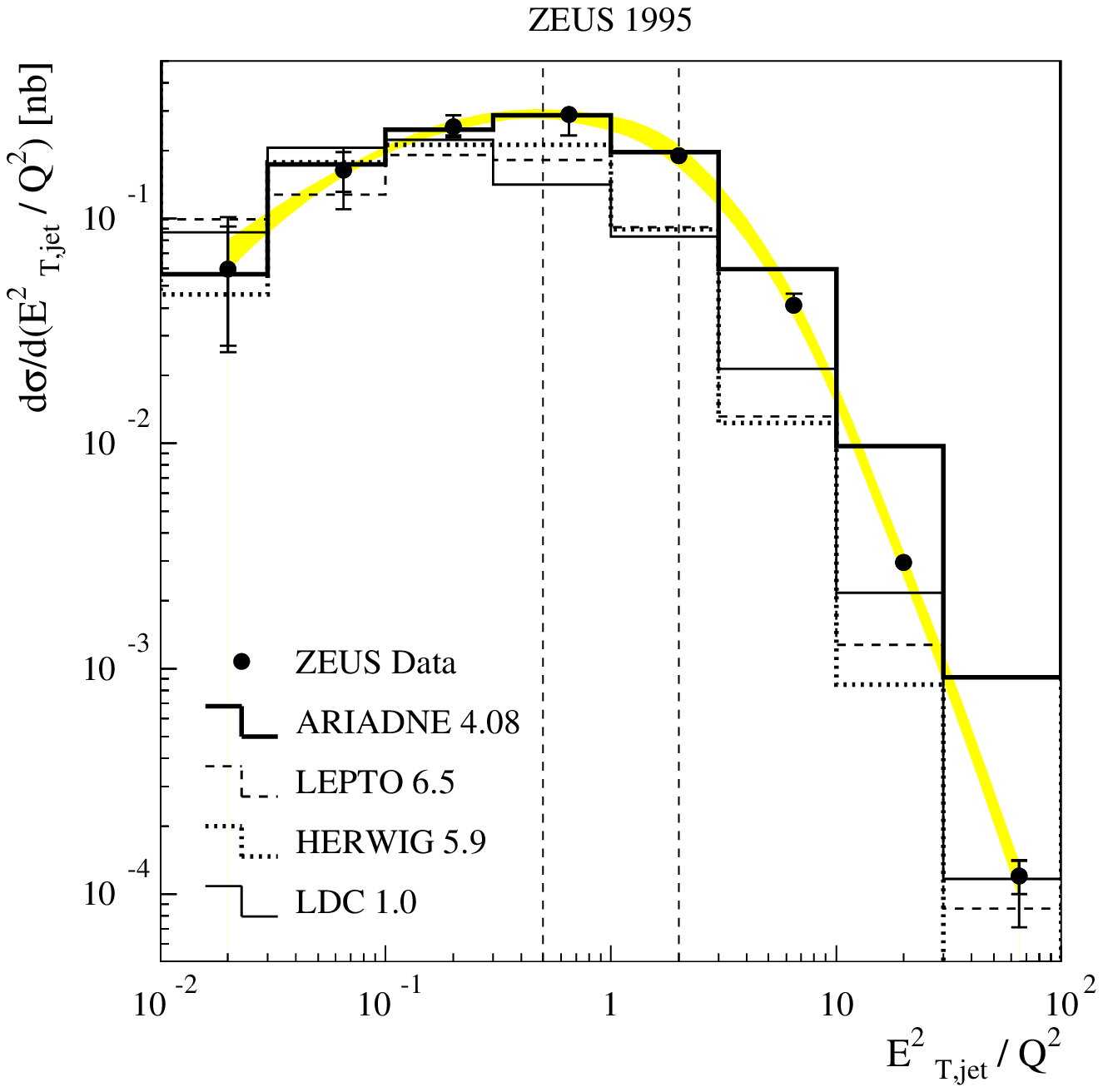,width=0.77\linewidth,clip=}
   \epsfig{file=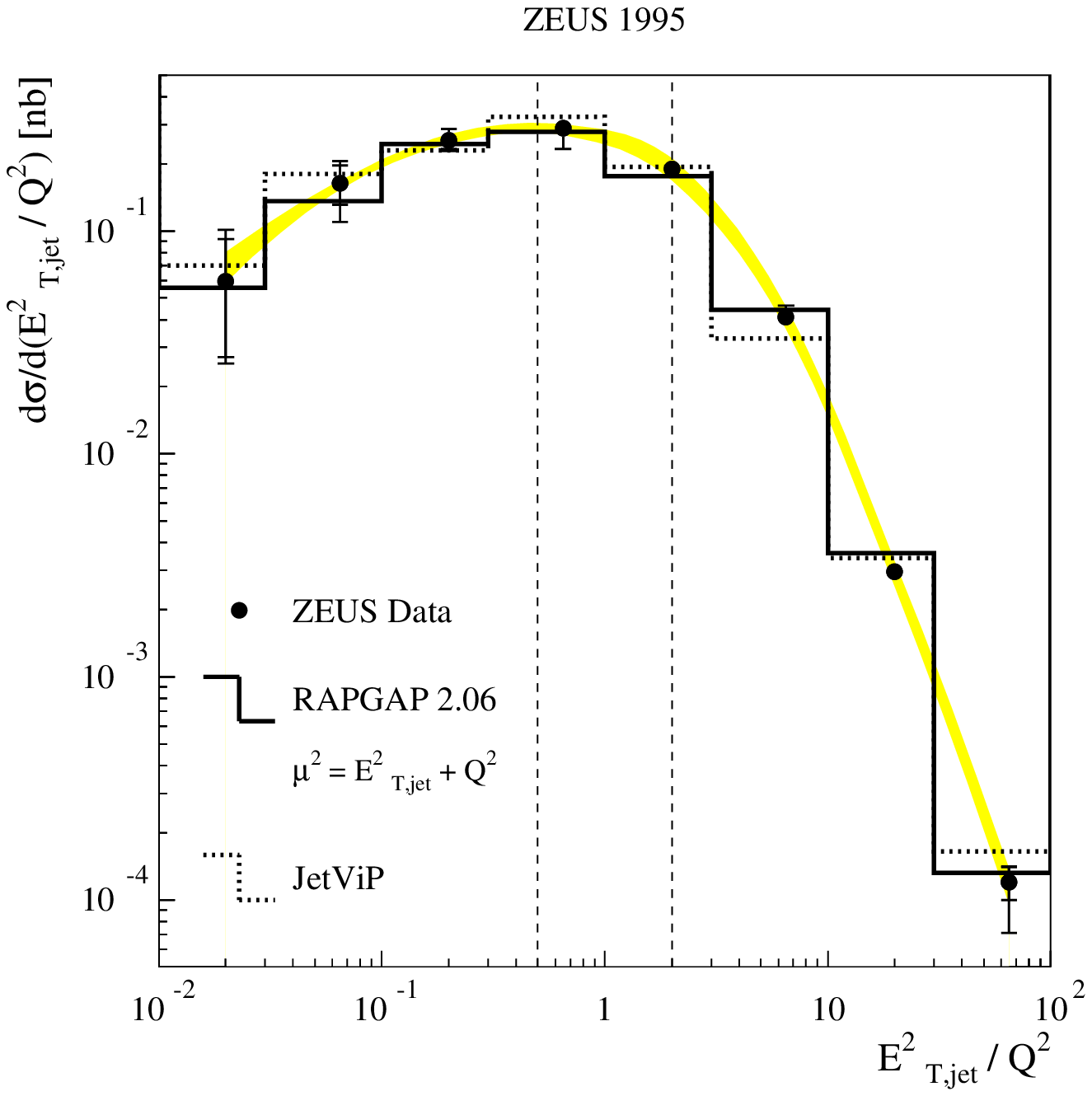,width=0.77\linewidth,clip=}
   \caption{\label{zeus.forwardjet} Cross-section for forward jet
   production from ZEUS\protect\cite{ZEUS.forward.jets} in comparison
   to calculations based on:  
       upper figure:  DGLAP (LEPTO and HERWIG) and colour-dipole model
   (ARIADNE); 
       lower figure: resolved virtual photons in LO (RAPGAP) and NLO
   (Jetvip).}  
  \end{center}
\end{figure}
Fig.~\ref{zeus.forwardjet} shows the cross-section for forward jets as a
   function of $E_T^2/\qsq$ from ZEUS\cite{ZEUS.forward.jets}. 
   None of the DGLAP based
   calculations are
   able to explain the data everywhere.
Calculations based on structure functions for {\it
   virtual photons} have been available since several years, where for
   the case 
   $E_T^2\gg\qsq$ the $\gamma^*$ is assumed to be resolved by the $E_T$
   of the jets. Starting from the highest $E_T$ somewhere along the
   ladder, two parton cascades are then evolved towards the photon
   and the proton. As this corresponds to $non-k_T$ ordering, it may be
   viewed as an approximation to new QCD dynamics such as those predicted
   by the BFKL or CCFM equations. Calculations based on resolved
   virtual photons are indeed able to explain the 
   data both in the DGLAP regime at $E_T\ll \qsq$ or $E_T\gg \qsq$ and
   in the BFKL regime at $E_T\approx \qsq$.
Apart from these data only weak evidence for BFKL type dynamics exists
   up to now \cite{bfkl.theory,bfkl.experimental.evidence}.

\begin{figure}[h]
   \epsfig{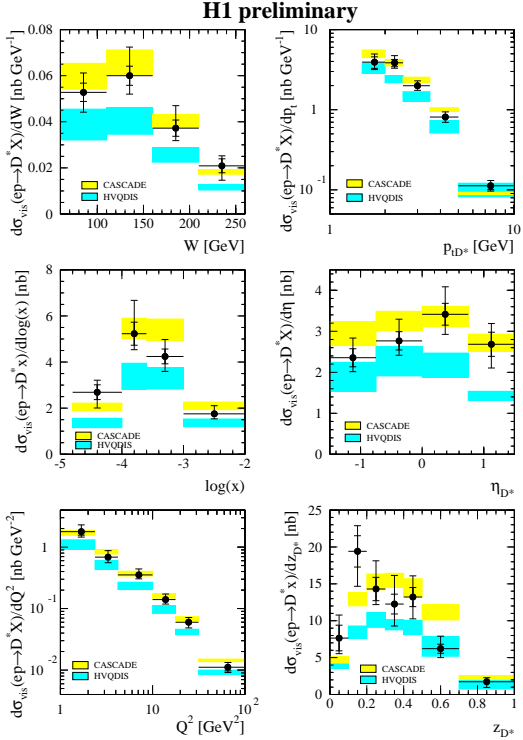}
   \caption{\label{h1.f2charm} Cross-section for $D^*$
   mesons from H1\protect\cite{H1.charm} as a function of $W, x$ and
   \qsq together with the 
   transverse momentum $P_{T,D^*}$, pseudorapidity $\eta_{D^*}$ and
   momentum fraction $z$ of the $D^*$. HVQDIS and CASCADE are NLO DGLAP
   and CCFM based calculations, respectively.}
\end{figure}

Similar to the forward 
   jet case, deep inelastic production of charm is a
   two-scale process and thus a test-case for effects beyond $k_T$
   ordering. Fig.~\ref{h1.f2charm} shows the charm production
   cross-section in    comparison 
   to DGLAP and CCFM based calculations. Note that the calculation 
   makes use of unintegrated parton densities 
   $f(x,k_T^2, \qsq)$ which are obtained from a fit to the H1 $F_2$
   data\cite{jung.ccfm}.
   While deviations from the data are 
   present in both approaches, the CCFM calculation does 
   better especially in the forward direction at large pseudorapidities
   $\eta$ close to the proton.
   This test of CCFM dynamics looks promising also 
   in details of the final state. 
\section{Diffraction} 
\begin{figure}[t]
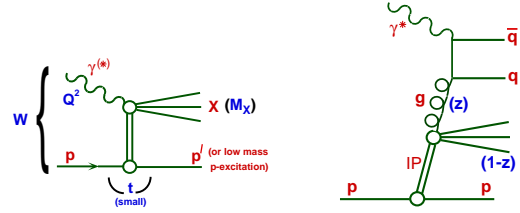

   \begin{center}
   \epsfig{file=newman_gendis.epsf,width=0.5\linewidth}
    \hspace*{0.1\linewidth}
   \epsfig{file=newman_bgf2.epsf,width=0.35\linewidth}
   \caption{\label{feyn.diffraction.inclusive} Feynman diagrams for
   diffractive scattering in $\gamma^*p$ collisions at HERA; inclusive
   scattering (left) and di-jet production (right).}
   \end{center}
\end{figure}
Processes in which a proton is scattered elastically are a
challenge to pQCD calculations as they must proceed via a colour
singlet exchange, in contrast to the standard approximation of
single quark or gluon exchange.
These processes are of fundamental interest 
as, in the end, they address the nature of colour confinement in
QCD\cite{confinement.bartels}. 

From {\it soft} hadronic processes it is 
known that beyond the expected exchange of
photons and mesons, an additional component must be present, which can
not be associated to any known particle\footnote{In the framework of
Regge theory\cite{regge} this exchange was labelled ``pomeron'' ($\pom$).}.
This colour singlet exchange has generally been assumed to be dominated by
gluons, however its precise nature  remained unclear.

The interest in diffraction was 
renewed when {\it hard} diffractive processes were observed in \ppbar
collisions\cite{ppbar.hard.diffraction} and $ep$ collisions
\cite{ep.hard.diffraction}, in which 
the partonic structure of the colour singlet exchange can be resolved.
\subsection*{Hard Diffraction at HERA}
Fig.~\ref{feyn.diffraction.inclusive} shows the diagram for deep
inelastic, inclusive diffractive scattering. 
\begin{figure}[h]
   \epsfig{file=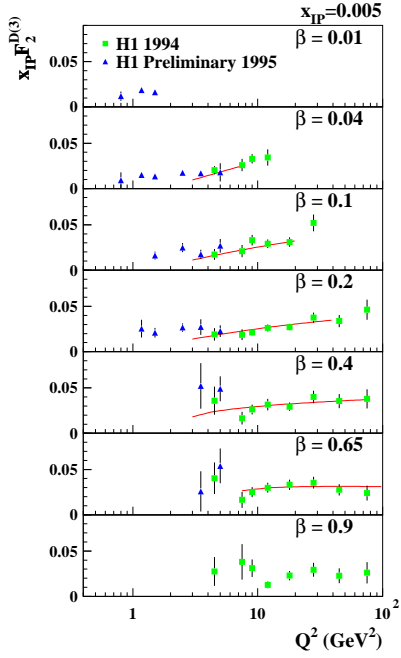,width=0.8\linewidth,clip=}
   \caption{\label{f2d3} The diffractive structure function $F_2^{D(3)}$
   from H1\protect\cite{H1.f2d3,H1.difflowq2}.}
\end{figure}
\begin{figure}[bthp] 
   \epsfig{file=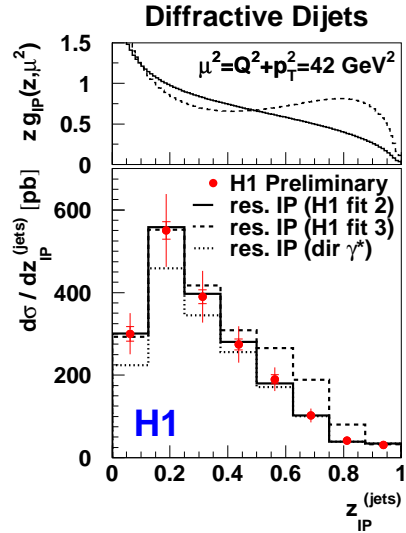,width=0.8\linewidth,clip=}
   \caption{\label{H1.diffractive.jets.inclusive} 
    Diffractive di-jet cross-section from
   H1\protect\cite{H1.diffractive.dijet} (bottom) in
   comparison to 
   predictions based on gluon densities (top) extracted by two QCD
   fits to the H1  $F_2^{D(3)}$ data. Here $z_{\pom}$ corresponds to
   $\beta$ for inclusive diffractive scattering.}
\end{figure}
\begin{figure}[bthp]
   \epsfig{file=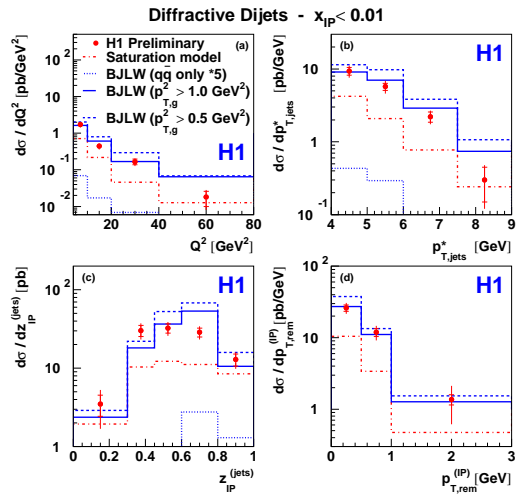,width=\linewidth}
   \caption{\label{h1.diffractive.jets.bartels} Diffractive di-jet
   cross-section from H1\protect\cite{H1.diffractive.dijet} in
   comparison to calculations (BJLW) based on the 
   exchange of 2 gluons which interact dominantly with a $q\bar{q}g$
   fluctuation of the $\gamma$.}
\end{figure}
\begin{figure}[bthp]
   \epsfig{file=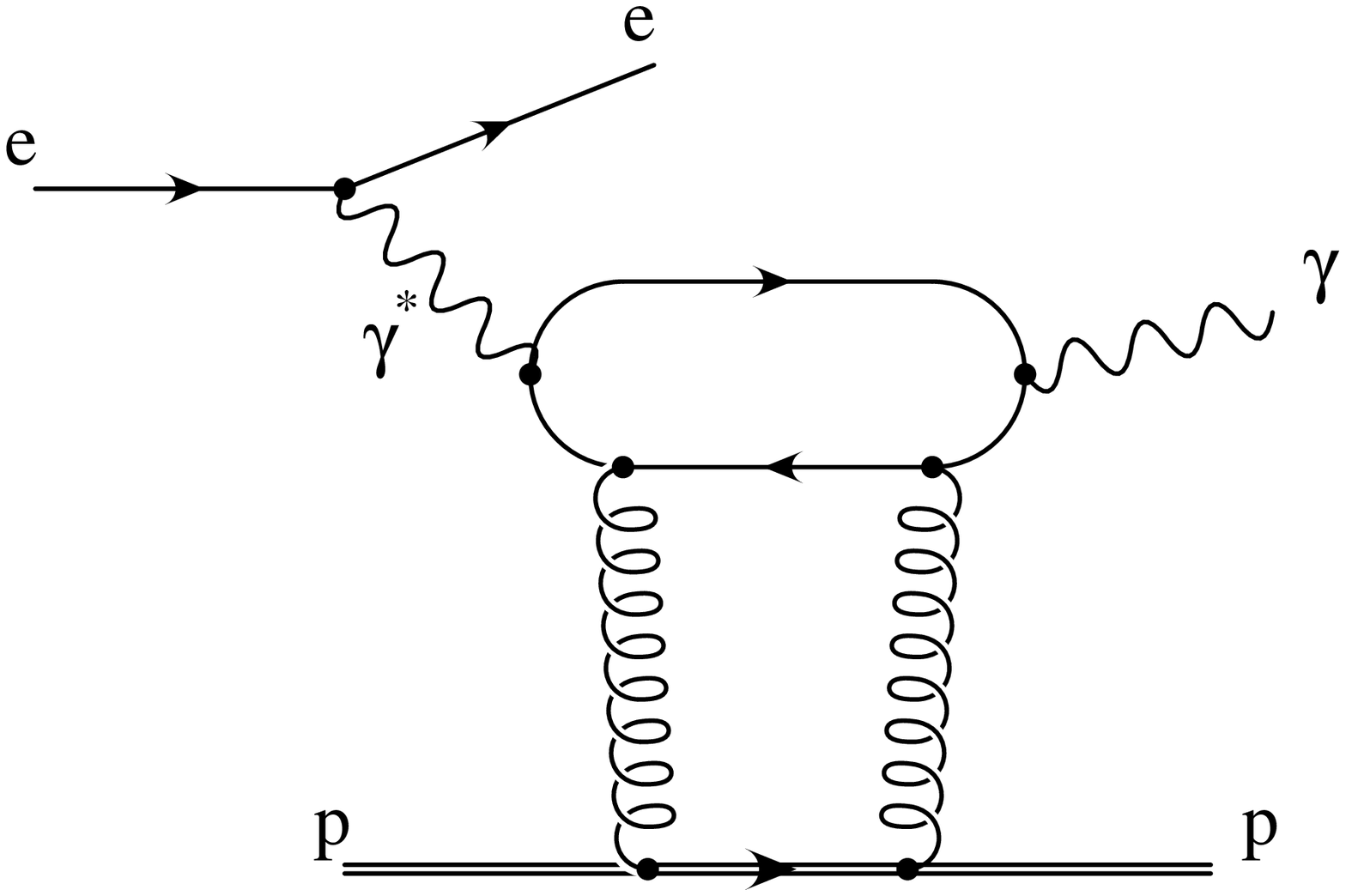,width=0.45\linewidth}
   \epsfig{file=H1prelim-00-017.fig6-col.epsi,width=0.5\linewidth}
   \caption{\label{h1.dvcs.cross} Feynman diagram and cross-section
   for Deeply Virtual Compton Scattering (DVCS) from
   H1\protect\cite{H1.dvcs}. The process 
   interferes with the Bethe-Heitler (BH) process $ep\ar e\gamma p$
   (where the $\gamma$ is emitted from the electron line)
   for which the contribution is shown separately.}
\end{figure}
The corresponding cross-section depends on four kinematic quantities:
$x$, \qsq, the momentum transfer squared $t$ at the proton vertex and
the momentum fraction \xpom of the colour singlet relative to the
proton. 

Important theoretical progress was recently achieved by the proof
of QCD hard scattering factorisation \cite{diffractive.factorisation}
for diffractive $ep$ scattering\footnote
{Note that 
factorisation is expected to break down in \ppbar collisions due to
secondary interactions between the two protons.},  
which states that the cross-section factorises into hard
partonic cross-sections and universal diffractive parton distributions 
\fdfour. The latter should evolve (at fixed \xpom and $t$) according
to the DGLAP evolution equations.
This proof puts diffraction on a solid basis for a treatment in pQCD
and experimental verification is highly desirable.

For inclusive diffractive scattering the HERA experiments 
usually integrate over $t$ and present their  
result as a structure function \fdthree (Fig.~\ref{f2d3}),
where $\beta=x/\xpom$ corresponds to the momentum fraction of the struck
parton in the colour singlet exchange.
The data\cite{H1.f2d3} show positive scaling violations up to large
$\beta$, indicating the dominance of 
gluons in the diffractive exchange.

Assuming in addition Regge factorisation\cite{regge}, 
$F_2^{D(3)}$ can be 
written as a flux factor for the colour singlet 
$f_{(\xpom,t)}$ times a structure function for the exchange $\fdtwo$.
QCD fits  of the scaling violation in \fdtwo yielded gluon
densities\cite{H1.f2d3} as shown in
Fig.~\ref{H1.diffractive.jets.inclusive} (top).  

A direct measure of the gluon distribution can be obtained from di-jet
production in  diffraction
(Fig. \ref{feyn.diffraction.inclusive}). The new data from
H1\cite{H1.diffractive.dijet}   (Fig.~\ref{H1.diffractive.jets.inclusive})
are compatible with both 
Regge and QCD factorisation and  nicely confirm the previous
$F_2^{D(3)}$ analysis which assumed Regge
factorisation.
 A new, QCD based calculation of this diffractive cross-section,
 which assumes dominant exchange
 of two gluons interacting with the $q\bar{q}(g) $ system emitted
 by the virtual photon (c.f. Fig. \ref{feyn.diffraction.inclusive}),
 leads to a reasonable description of the data at small \xpom
 (Fig.~\ref{h1.diffractive.jets.bartels}). 
Again unintegrated gluon densities are  employed here.
Similar calculations have also become available for the diffractive
production of vector-mesons (Fig.~\ref{vm.all}) and for 
Deeply Virtual Compton Scattering $\gamma^* p\ar \gamma p$ (DVCS)
(Fig.~\ref{h1.dvcs.cross}).  
It is remarkable that perturbative calculations are now able, with only
a few free parameters, to describe a number of hard diffractive
processes in $ep$ scattering. It would be of high interest to
complement these data with measurements at high
$t$ where the colour-singlet itself might be calculable in pQCD.
\subsection*{Hard Diffraction at the Tevatron}
Both D0 and CDF have investigated processes where jets
with large $E_T$ are employed to study the partonic structure of the
diffractive exchange. The most striking observation is that
the overall rate of diffractive processes is much smaller (by
factors 5 to 20 \cite{D0.diffraction}) in comparison to the findings
at HERA, where they 
contribute as much as 10\% to the DIS cross-section.

\begin{figure}[bthp]
   \epsfig{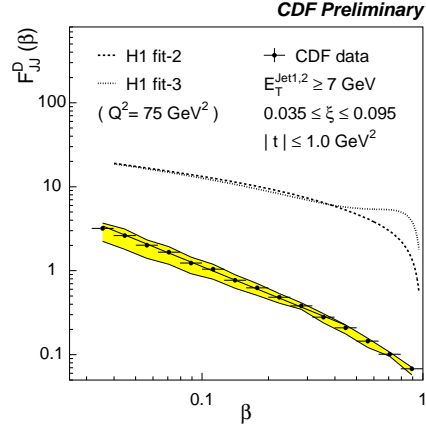}
   \caption{\label{cdf.diffractive.jets.hera} The diffractive
   structure function extracted  from jet data by
   CDF\protect\cite{CDF.diffractive.jets} in comparison to an 
   expectation based on the H1 diffractive parton densities.}
\end{figure}
\begin{figure}[bthp]
   \epsfig{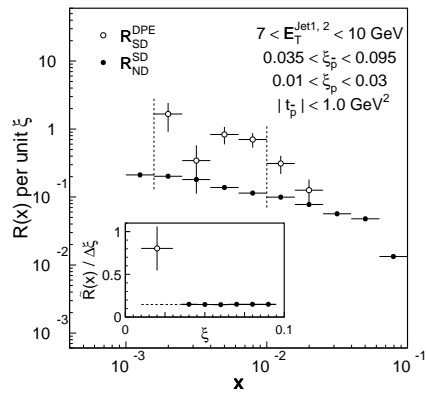}
   \caption{\label{cdf.diffractive.ratios} Ratios of double to single
   diffractive cross-sections $R_{DPE}^{SD}$ and single to
   non-diffractive  cross-sections $R_{SD}^{ND}$ from CDF\protect\cite{CDF.diffractive.ratios}.}
\end{figure}

Fig.~\ref{cdf.diffractive.jets.hera} shows the effective structure
function from jets $F_{jj}^D$ from CDF for events where the
elastically scattered proton was tagged at very small scattering
angles. 
The data are far below a calculation which is based on the H1
parton densities extracted from $F_2^{D(3)}$
(Fig.~\ref{H1.diffractive.jets.inclusive}).  
QCD factorisation obviously is badly broken for diffractive \ppbar 
collisions. The same conclusion is obtained using only Tevatron data
from the ratios of double- to single and single to non-diffractive
cross-sections, which are significantly different
(Fig.~\ref{cdf.diffractive.ratios}).
It is noted that the QCD factorisation proof for $ep$
scattering does not apply to \ppbar collisions.
The large factorisation breaking observed when extrapolating from HERA
results to Tevatron processes reflects that a point-like virtual
photon (or a $q\bar{q}$ fluctuation of small transverse size) is able
to pass through the proton without destroying it, whereas two large
 proton remnants will destroy each other in most cases.
While this prohibits at present an interpretation of the \ppbar data in terms
of universal diffractive parton densities, the mechanism of factorisation
breaking is in itself of interest and a challenge for the
understanding of diffraction in hadron-hadron
collisions\cite{diffraction.hera.tevatron}. 
\section{Conclusion}
Among the most active fields of QCD are the regions
where perturbative approaches are
difficult, namely at large distances close to the confinement limit,
at high energies or low $x$, in regions of large parton densities
and where multi-parton exchange becomes crucial.
The entire field is driven by the availability of very precise
data which are needed for detailed tests of QCD.

At low $x\gtrsim 10^{-3}$ the structure function data from HERA constrain
the gluon density with a precision of better than 3\% at high \qsq,
which paves the way for  
significant tests of the Higgs sector at the Tevatron/LHC. While
conventional DGLAP evolution describes the inclusive data down to
$\qsq\gtrsim 1\gev$,  more exclusive measurements of the hadronic
final state indicate the need for calculations beyond strong $k_T$
ordering. Evidence for BFKL effects from HERA, LEP or Tevatron are still
weak, in spite of the fact that theoretical uncertainties seem to be
better controlled\cite{bfkl.theory}. The first CCFM calculations based
on parton densities unintegrated in $k_T$ look promising when compared
with HERA data on charm production.

The LEP and HERA experiments have provided data on the photon
structure with much improved precision. As existing photon parton
densities seem not to be sufficient to describe all measurements a new effort
in the understanding of the photon structure in NLO QCD is required.

In diffraction the comparison between hard inclusive and exclusive
 processes at HERA has led to a consistent picture of the structure of
 diffractive colour singlet exchange which is dominated by gluons. 
A challenge here is the application of the HERA results to hadron
 collisions, where QCD factorisation in diffraction is shown to be
 broken.
\section*{Acknowledgements}
It is a pleasure to thank the organizers, Y.Yamazaki and the technical
staff in Osaka 
for their help, K. Borras, J. Dainton, E. Elsen, B. Foster, M. Klein, 
P. Newman and R. Nisius for 
useful comments to the manuscript and many others for 
valuable discussions.


\begin{thebibliography}{99}
\bibitem{nania} R. Nania, \these.
\bibitem{pellegrino} A. Pellegrino, \these; also \protect\cite{h1.f2}.
\bibitem{h1.f2} M. Klein, proceedings Lepton-Photon 1999, SLAC.
\bibitem{h1.qcd} F. Zomer, \these. 
\bibitem{first.hera.f2} \collh \nuclphys{B407}{1993}{515}; 
                        \collz \physlet{B316}{1993}{412}. 
\bibitem{mellado} B. Mellado, \these.
\bibitem{dipolefac} B.L.Ioffe, \physlet{30B}{1969}{123}.
\bibitem{dipolemod} 
          K.Golec-Biernat, M.W\"{u}sthoff, \\ \physrev{D59}{1999}{014017};\\
                 M.McDermott, hep-ph/9912547, and references therein.

\bibitem{lhc} S. Catani et. al., hep-ph/0005025, hep-ph/0005114.
\bibitem{lep.higgs} A.Gurtu, \these; P. Igo-Kemenes, \these, and
talk at the LEPC Nov.3rd, 2000.
\bibitem{DGLAP} 
  Y. Dokshitzer, \journal{Sov. Phys. JETP}{46}{1977}{641};
  V. Gribov, L. Lipatov, \journal{Sov. J. Nucl. Phys.}{15}{1972}{438 and 675};
  G. Altarelli, G. Parisi, \nuclphys{B126}{1977}{298}.
\bibitem{NNLO} W.L. van Neerven, A. Vogt, hep-ph/0006154;
               A.D. Martin et. al., hep-ph/0007099.

\bibitem{BFKL} 
   E.Kuraev, L.Lipatov, V.Fadin, \journal{Sov. Phys. JETP}{45}{1977}{199}; 
   Y. Balittski, L. Lipatov, \journal{Sov. J. Nucl. Phys}{28}{1978}{822};
   L. Lipatov, \journal{Sov. Phys. JETP}{63}{1986}{904}.
\bibitem{CCFM} M.Ciafaloni, \nuclphys{B296}{1988}{49};
        S. Catani, F. Fiorani, G. Marchesini,
                      \physlet{B234}{1990}{339},\nuclphys{B336}{1990}{18};  
        G. Marchesini,\nuclphys{B445}{1995}{45}.
\bibitem{jpsi.qcd} M. Ryskin et. al., \journal{Z. Phys.}{C76}{1997}{231};
                   L. Frankfurt et. al., \physrev{D57}{1998}{512}.
\bibitem{valencegluon.h1} \collh contr. paper ICHEP 1997, Jerusalem.
\bibitem{valencegluon.zeus} \collz \epj{C7}{1999}{609}.
\bibitem{foster} B. Foster, hep-ex/0008069, and references therein.
\bibitem{lowx.exclusive} U. Maor, \these. 
\bibitem{levonian} S. Levonian, \these.  
\bibitem{nisius} R. Nisius, \journal{Phys.Rept.}{332}{2000}{165}.
\bibitem{erdmann} M. Erdmann, Springer Tracts in Modern Physics 138
(1996), DESY-96-090.
\bibitem{nisiuspriv} R. Nisius, private communication.
\bibitem{lep.photon.x} \collo, CERN-EP-2000-82.
\bibitem{lep.charm} A. B\"{o}hrer, \these.
\bibitem{opal.f2charm} \collo \epj{C16}{2000}{579}.
\bibitem{hera.b} \collh \physlet{B467}{1999}{156}.
\bibitem{H1.jets.gammap} \collh \physlet{B483}{2000}{36}.
\bibitem{Zeus.jets.gammap} \collz \epj{C11}{1999}{35}.
\bibitem{wengler} Th. Wengler, \these. 
\bibitem{bfkl.theory} L.H. Orr, \these.
\bibitem{bfkl.experimental.evidence} \collh \physlet{B462}{1999}{440};
       \colld B.Pope, \these; \colll M.Wadhwa, \these.
\bibitem{advocate.forward.jets} 
   A. Mueller, \journal{Nucl.Phys. (Proc.Suppl.)}{18C}{1991}{125},
               \journal{J.Phys.}{G17}{1991}{1443}; 
   J. Kwiecinski, \epj{C9}{1999}{611}. 
\bibitem{ZEUS.forward.jets} \collz \physlet{B474}{2000}{223}.
\bibitem{jung.ccfm} H. Jung, DIS workshop 1999, hep-ph/9905554;
      S.Baranow, H.Jung, N. Zotov, hep-ph/9910210.
\bibitem{H1.charm} E. Tzamariudaki, \these.
\bibitem{regge} P.D.B. Collins, {\it An Introduction to Regge Theory
and High-Energy Physics}, Cambridge 1977.
\bibitem{confinement.bartels} J.D.Bjorken, hep-ph/0008048; 
        J.Bartels, H.Kowalski, hep-ph/0010345.
\bibitem{ppbar.hard.diffraction} \coll{UA8} \physlet{B211}{1988}{239};
                                            \physlet{B297}{1992}{417};
                                            \physlet{B421}{1998}{395}.
\bibitem{ep.hard.diffraction} \collz \physlet{B315}{1993}{481}; 
                              \collh \nuclphys{B429}{1994}{477}.
\bibitem{diffractive.factorisation} 
          J. Collins, \physrev{D57}{1998}{3051}, erratum.
\bibitem{H1.f2d3} \collh \journal{Z. Phys.}{C76}{1997}{613}. 
\bibitem{H1.difflowq2} \collh contr. paper to ICHEP 1998, Vancouver.
\bibitem{H1.diffractive.dijet} \collh F.P. Schilling, proceedings DIS 2000, Liverpool.
\bibitem{zeus.dvcs} \collz contr. paper to EPS99, Tampere.
\bibitem{H1.dvcs} \collh R. Stamen, proceedings DIS 2000, Liverpool;
L. Favart, \these.
\bibitem{D0.diffraction} A. Sznadjer, \these.
\bibitem{CDF.diffractive.jets} \collc  \prl{84}{2000}{5043}.
\bibitem{CDF.diffractive.ratios} \collc \prl{85}{2000}{4215}.
\bibitem{diffraction.hera.tevatron} 
       F. Hautmann, D.E. Soper,\physrev{D63}{2000}{011501}; 
       V. Khoze, A. Martin, M. Ryskin, hep-ph/0007083; and references therein. 
\end{thebibliography}
\end{document}